\documentclass[conference]{IEEEtran}
\usepackage{graphicx}
\usepackage{subfig}
\usepackage{rotating}
\usepackage{caption}
\ifCLASSINFOpdf
\else
\fi
\begin{document}
%
\title{Automated Document Indexing via Intelligent Hierarchical Clustering: A Novel Approach}

\author{\IEEEauthorblockN{Rajendra Kumar Roul}
\IEEEauthorblockA{Department of Computer Science\\
BITS-Pilani K.K. Birla Goa Campus\\
Zuarinagar, Goa, India - 403726\\
Email: rkroul@goa.bits-pilani.ac.in}
\and
\IEEEauthorblockN{Shubham Rohan Asthana}
\IEEEauthorblockA{Department of Computer Science\\
BITS-Pilani K.K. Birla Goa Campus\\
Zuarinagar, Goa, India - 403726\\
Email: asthana.st.francis@gmail.com}
\and
\IEEEauthorblockN{Sanjay Kumar Sahay}
\IEEEauthorblockA{Department of Computer Science\\
BITS-Pilani K.K. Birla Goa Campus\\
Zuarinagar, Goa, India - 403726\\
Email: ssahay@goa.bits-pilani.ac.in }}

%


\maketitle

\begin{abstract}
With the rising quantity of textual data available in electronic format, the need to organize it become a highly challenging task. In the present paper, we explore a document organization framework that exploits an intelligent hierarchical clustering algorithm to generate an index over a set of documents. The framework has been designed to be scalable and accurate even with large corpora. The advantage of the proposed algorithm lies in the need for minimal inputs, with much of the hierarchy attributes being decided in an automated manner using statistical methods. The use of topic modeling in a pre-processing stage ensures robustness to a range of variations in the input data. For experimental work 20-Newsgroups dataset has been used. The F-measure of the proposed approach has been compared with the traditional K-Means and K-Medoids clustering algorithms. Test results demonstrate the applicability, efficiency and effectiveness of our proposed approach.  After extensive experimentation, we conclude that the framework shows promise for further research and specialized commercial applications.
\end{abstract}


%
\IEEEpeerreviewmaketitle

\section{Introduction}

With the unprecedented rise in the number of knowledge sources - especially from the Internet, the amount of textual data available to any end user has become pretty vast. Internet size is at least 4.62 billion pages.\footnote{www.worldwidewebsize.com} According to the latest survey\footnote{http://www.factshunt.com/2014/01/world-wide-internet-usage-facts-and.html}, approximately 4.354 billion people are using Internet actively out of 7.1 entire human population. As a result, ways to organize this data and retrieve information from it in an efficient and accurate manner is the need of the hour. Document clustering, the primary domain of research tackled in this paper, is usually performed for various reasons pertaining to information retrieval – including document organization, summarization and classification. Our work focuses on using document clustering to build an `index' of sorts - to organize a given set of documents into an intelligent hierarchy of collections and sub-collections, similar to a tree structure.

Building a hierarchical structure for document organization/indexing has various advantages. For starters, it aids human understanding of the entire document collection as a whole, facilitating a user's browsing of the large amount of data without having to wade through irrelevant information. Secondly, it also helps automated systems with efficient information retrieval pertaining to specific user queries. In a cluster hierarchy, each parent-child association is analogous to a topic-subtopic relationship. This means that all documents belonging to a sub-cluster together denote a subtopic of the overall subject pertaining to its parent cluster(which includes the documents assigned to each of its sub-clusters). As a result, if the cluster hierarchy output by our framework is implemented as a directory system, then a user would be able to navigate folder-wise to only that end-folder which contains the documents that are most relevant to him – provided our work is augmented with some labeling system to label every cluster according to its content.

	However, there are a lot of issues that need to be dealt with while trying to accomplish clustering of documents – including scalability and accuracy. The most basic one of them is the curse of dimensionality. In most primitive methods, the documents are represented as vectors using bag-of-words technique, complemented by improvements like TF-IDF \cite{aizawa2003information}. However, the space of dimensions equal to the size of the vocabulary introduces great noise, reducing accuracy and increasing time complexity substantially. It also does not deal with problems like synonymy and polysemy. To tackle these problems, we use topic modeling in the form of Latent Semantic Indexing(LSI)\cite{a5} to reduce documents to the considerably less noisy and more information-rich `semantic space'. Moreover, components of the actual clustering algorithm have also been optimized with inspiration from methods such as Principle Direction Divisive Partitioning(PDDP)\cite{a16}.

	One of the major advantages of the framework proposed in this paper is its ability to intelligently judge the attributes of the cluster hierarchy. As a result, it does not require the user to have extensive domain knowledge about the text contained in the documents. For example, most hierarchical clustering techniques existing in current literature have the user input as the number of child nodes that each node in the cluster tree will have. Not only this quite rigid conceptually, but also it bound to decrease the accuracy of information retrieval. We use a unique flat clustering algorithm which intelligently determines the number of sub-clusters for each cluster in the hierarchy should have. Moreover, a simple tree-based algorithm enables easy navigation of the entire `index' for an automated information retrieval system. We compared the F-measure of our approach with K-Means \cite{a1} and K-Medoids \cite{a2} algorithms to measure the system performance.

	The rest of the paper is organized as follows - Section 2 sketches the related work in existing literature; Section 3 describes our framework in detail; Section 4 showcases the pertinent experimental results and Section 5 concludes with remarks about future work possible. 
	
\section {Related Work}
\nocite{a12}
Document clustering has received notable attention as a problem of research in the field of information retrieval. The most basic methods to achieve this involve running is the classic clustering algorithms like K-Means and K-Medoids over a set of documents represented as bag-of-words. More recently, researchers have started adopting means of representing documents more intelligently, in reduced dimensional spaces\cite{a7}.

	Use of topic modeling in text corpus clustering can be broadly classified into two categories: as a dimensionality reduction method, or as a method of direct probabilistic clustering. Probabilistic methods like Latent Dirichlet Allocation (LDA)\cite{a3} and Probabilistic Latent Semantic Analysis (PLSA)\cite{a4} have been investigated as soft clustering techniques by various researchers. Usually, frameworks involving these techniques model the corpus with a pre-defined number of topics, where each topic is taken to be a cluster. The output suggests the probability of any given document belonging to any one of the computed topics.

	Another notable set of methods for document clustering involve matrix factorization, like Latent Semantic Indexing, Non-negative Matrix Factorization(NMF)\cite{a6} and Concept Factorization\cite{a10}. Some stochastic search techniques have also been coupled with these algorithms to provide good results in document clustering - for example coupling genetic algorithms with LSI\cite{a15} or using LSI with Particle Swarm Optimization (PSO)\cite{a8}. Some works also use heuristic search with LSI to cluster text documents, like \cite{a9}. As a dimensionality reduction algorithm, LSI can be used along with K-Means to achieve very good results for large document sets.

	Most of the aforementioned techniques achieve flat clustering, which basically means a non-hierarchical or `one-level up' clustering. As for hierarchical methods, the most universally used technique is agglomerative hierarchical clustering. This paradigm builds a hierarchy bottom-up by iteratively computing the similarity between the current set of clusters, and merging the two most similar ones. Hence, agglomerative clustering assumes only two sub-clusters per cluster, and suffers from a large time complexity in case of a huge corpus.

	As for divisive clustering, K-Means and Bisecting K-Means \cite{b2} are mostly used for partitioning. Partitioning with K-Means continues until you reach one-document-per-cluster, or until a stopping criterion is met. The stopping criterion most widely used in literature is the Bayesian Information Criterion(BIC)\cite{a11}. Bisecting K-Means works by dividing clusters into two parts until required number of clusters is reached. Bisecting K-Means is more efficient and accurate than K-Means as well as agglomerative clustering. But lack of knowledge about the number of clusters required is a common problem with this method. In our framework, we use a divisive paradigm inspired by Bisecting K-Means, which splits the set of documents into an optimal number of clusters automatically. This is then repeated bottom-up level by level, until the 'root' of the cluster tree is obtained.

	Finally, a unique keyword-based algorithm, Frequent-Itemset hierarchical clustering has also been developed \cite{a14}. It is able to build a non-conceptually-rigid hierarchy of document clusters, given a corpus. However, since the method is entirely dependent on keywords, the problems of synonymy and polysemy remain unanswered.
	
\section{Document Indexing Framework}

This section describes the methodology behind the proposed document indexing structure.

\subsection{Topic Modeling using Latent Semantic Indexing}

Before getting to the clustering framework, topic modeling is performed on the given set of documents - assuming standard pre-processing steps such as stopword removal and stemming have already been performed on the dataset which transforms the documents from being vectors in keyword-space, to vectors in semantic space. Topic modeling achieves the two-fold job of preventing the curse of dimensionality and at the same time discovering the `abstract topics' present in the document collection. We achieve this using LSI, an algorithm that uses Singular Value Decomposition(SVD) to extract the `concepts' present in the text corpus.

	We use LSI over the more popular Latent Dirichlet Allocation (LDA) method. This is chiefly because the clustering part of our framework interprets the document vectors as data points in space, rather than as a collection of numbers denoting probabilities/membership. In other words, we essentially use LSI as a dimensionality reduction technique. The number of LSI topics is decided empirically, usually via limited previous knowledge about the text content or by keeping the number of topics proportional to the size of document collection.

	The vectors output by the LSI algorithm are then passed on to the document clustering framework, described in the subsequent subsection.

\subsection{The Hierarchical Document Clustering Algorithm}

\subsubsection{Flat Clustering using Intelligent Divisive Partitioning}

The crux of our document clustering methodology is a divisive-partitioning but flat-clustering paradigm. Divisive partitioning starts off with a `super-cluster' of all known data points. It then recursively partitions each cluster formed on the way, until either a stopping criterion is met, or one left with one document per cluster at the `leaves'. Though divisive partitioning itself traditionally produces a hierarchy of clusters, the number of sub-clusters per cluster is hard-coded or input by the user. As a result, the cluster hierarchy generated is not completely suited to the patterns shown by the data. We use divisive partitioning as a flat-clustering algorithm by considering only the leaf nodes/clusters with respect to the partitioning `tree'.

	A `complete' partitioning tree would mean the leaves having one document each, as previously mentioned. In our approach, we prune this tree at appropriate places and avoid the further division of some sub-clusters. This is achieved by using a custom stopping criterion. Thus, if a cluster that is generated midway during the algorithm's run meets this criterion, it is not partitioned any further.

	We use a multivariate Gaussian distribution\footnote{http://cs229.stanford.edu/section/gaussians.pdf} to model every cluster of documents. The centroid  ($\mu$) and covariance matrix ($\Sigma$) with respect to each cluster $C$ can be obtained as follows: 

\begin{equation}
\mu = \frac{\sum_{x \epsilon C}{x}}{n_C}
\end{equation}

and

\begin{equation}
\Sigma_{i,j} = cov(X_i, X_j)
\end{equation}

\noindent where $n_C$ denotes the number of elements in cluster $C$, $X_i$ denotes collection of the $i$th entries of all vectors in $C$ and $\Sigma_{i,j}$ denotes the $(i, j)$th element in the covariance matrix $\Sigma$.

	This distribution corresponding to a document cluster is then used to measure its `quality'. We compute the quality of a cluster, $Q_C$ using the following definition:

\begin{equation}
Q_C = \frac{1}{E[D_M(x_C, \mu_C, \Sigma_C)]}
\end{equation} 

\noindent where $D_M(x, \mu_C, \Sigma_C)$ denotes the Mahalanobis distance\cite{a13} of data point $x$ from the cluster $C$.

The mahalanobis distance of a data point from the centroid of a cluster, with respect to the cluster's covariance matrix, is given as follows:

\begin{equation}
E[D_M(x_C, \mu_C, \Sigma_C)] = \sqrt{(x - \mu_C)^{T}{\Sigma_C}^{-1}(x - \mu_C )}
\end{equation} 
	
	Compared to the Euclidean distance, the Mahalanobis distance is a better measure of the dissimilarity of a point with respect to the centroid, since it takes the covariance values (and hence the `shape' of the cluster) into account. We incorporate this measure of cluster quality into the stopping criterion for divisive partitioning. A cluster $C$ is not partitioned further if and only if the following criterion is met: 

\begin{equation}
Q_C \leq \frac{\sum_{C' \epsilon \Delta(C)} Q_C'}{|\Delta(C)|} 
\end{equation}

\noindent where $\Delta(C)$ denotes the set of clusters obtained after splitting cluster $C$.

	The splitting at each level of the divisive algorithm is performed using a standard K-Means approach, with the number of required clusters set to 2 (binary splitting). However, instead of using a random initialization of the two centroids, we use a method inspired by PDDP. By this method, the two centroids used in the initialization of the splitting step are given as follows:

\begin{equation}
\mu_1 = \mu_C + v_P * \frac{\sum_{x \epsilon C} x \ . \ v_P}{n_C}
\end{equation}

and

\begin{equation}
\mu_1 = \mu_C - v_P * \frac{\sum_{x \epsilon C} x \ . \ v_P}{n_C}
\end{equation}

\noindent where $v_P$ denotes the first principle direction computed over the data points in $C$ using Principle Component Analysis(PCA)\footnote{http://nyx-www.informatik.uni-bremen.de/664/1/smith\_tr\_02.pdf}. Such an initialization proves useful in increasing the accuracy of the procedure (as the final centroids tend to lie very near to the initialization pair), while at the same time reducing the convergence time substantially.

	Thus, any given cluster is broken down intelligently into an `ideal' number of sub-clusters by considering only the leaf-clusters obtained after a run of the above described algorithm.
	
\subsubsection{Bottom-Up Hierarchical Clustering}

We employ the technique described in the previous part of this subsection multiple times, to build a hierarchy of clusters in a bottom-up, level by level fashion.  Consider a cluster tree with levels of nodes marked from 0 (leaves) to $n-1$ (the root), where $n$ is the height of the tree. Level 0 denotes the bottom-most layer consisting only of documents. After running the flat clustering algorithm on this set of documents, we obtain the upper level of clusters at Level 1.

	It is to be noted that K-Means clustering, performed several times in the aforementioned flat clustering approach, promotes the development of circular-shaped clusters. Therefore, in a topic modeling scenario and also otherwise, the centroid of a cluster can be considered as the optimal `representative' for it. With this idea in mind, we gather the centroids of all the clusters of Level 1, and run the flat clustering algorithm on them – to obtain Level 2. This process goes on until, at a certain level, the set of centroids no longer splits into multiple clusters, thus giving us the top `super-cluster' – that is the root of the cluster hierarchy. It is interesting to note that some clusters in a certain level, say Level $i$, may contain only one cluster from Level $i-1$. In such a case, they are merged into one cluster to avoid unnecessary depth of the cluster tree.

	However, it must be remembered that as we go higher up the hierarchy to upper levels, the data-set under consideration becomes more and more sparse. Therefore,  suitable modifications must be made to the stopping criterion to deal with this change effectively. We do this by introducing a decay factor in the previously mentioned stopping criterion. The new condition for not splitting a cluster becomes - 

\begin{equation}
Q_C \leq \beta \ * \ \frac{\sum_{C' \epsilon \Delta(C)} Q_C'}{|\Delta(C)|} 
\end{equation}

\noindent where $\beta$ denotes the `decay factor'. The lesser the decay factor, the greater is the number of nodes in the higher level.

	Using this described approach, we are able to build a hierarchy of clusters for any given set of documents, which effectively represents an `index' over them. Using sophisticated keyword extraction techniques(such as noun as keyword), the top keywords for each node in the cluster tree can be noted and used for appropriate labeling. This will ensure that even human operators will be able to browse through the index (which can be implemented as a directory structure) in an effective manner.

\subsection{Categorization of Information Needs}

For efficient and fast information retrieval using the proposed framework, it might be necessary to classify a given `information-need' to some cluster of documents. This can be aided by exploiting the tree structure of the hierarchy. But before turning to the classification problem, it is necessary to represent the user's query in an appropriate manner. Such a query can traditionally be regarded as a bag of words. If query augmentation methods are used, it can be enriched to have a better vocabulary. Once the appropriate bag of words form is prepared, the query can be converted into a semantic-space vector using the LSI model that was previously generated for the original set of documents. This vector is then used for categorization to the appropriate cluster.

	First, the multivariate Gaussian distribution parameters of each node in the document cluster tree/hierarchy are calculated. This is purely a one-time job, and the same parameters can be used for every query that is input to the framework. To do this, we need to construct the document set pertaining to every node D. This can be achieved using the following recursive definition:
\\

\begin{verbatim}
Document-Set(node D):
	    If D is a leaf:
	        Document-Set(D) =
	         set of all documents present in D
	    Else:
	        Document-Set(Dn) = 
	          Union({Document-Set(Di) for each
	          child(sub-cluster) Di of D})
\end{verbatim}

Once the document set is constructed, the centroid and covariance matrix corresponding to each  cluster in the hierarchy can be calculated using (1) and (2). The Document-Sets pertaining to the cluster nodes can also be used for labeling them based on their content.

	A semantic-space LSI vector coming in can be categorized into the hierarchy using a simple tree-based algorithm as follows: 

\begin{verbatim}
Categorize(node D, query d):
    If D is a leaf:
       return D
    Else:
       D_next = that sub-cluster with 
               respect to whom mahalanobis 
               distance of query d is least
              (let the distance be d')
        If d' < Mahalanobis-dist(D, d):
            return Categorize(D_next, d)
        Else:
            return D
\end{verbatim}

	At the top level, this function will be called as `Categorize(root R, query q)'. The tree-search algorithm described above ensures that the set of documents most relevant to the query's information need are computed as efficiently as possible, without having to deal with all the documents present in the original corpus.
	
\section {Experiments and Results}

Experimentation pertaining to the proposed framework was conducted using the popular 20 Newsgroups dataset\footnote{http://qwone.com/$\sim$jason/20Newsgroups/}, mainly due to the size of the corpus and the human annotation of article topics. Each of the articles in the dataset is already classified to one of the 20 `newsgroups' originally collected by Ken Lang having 20 classes. Removing duplicates, the dataset contains 18,828 articles, out of which 11,293 were used by us for generating the document index, and the rest 7,535 were used for testing purposes. The code was executed on a supercomputer with 16 Processors - Intel Xeon(R) CPU E5-2680 0 @ 2.7 GHZ, 32GB RAM running Red Hat 4.4.5-6. In each of the test runs, the documents categorized to the same cluster show remarkable conceptual similarity. Moreover, the human annotation of the topics matches well with the framework's output, though the framework is able to find subtopics inside larger human-given topics too. The importance of topic modeling is seen in the fact that some articles from the topic `sci.electronics' were put in the cluster containing articles on `comp.sys.mac.hardware', since they talked about a common subject of computer-related electronics and hardware.
\begin{figure}
\centering
\includegraphics[width=10cm]{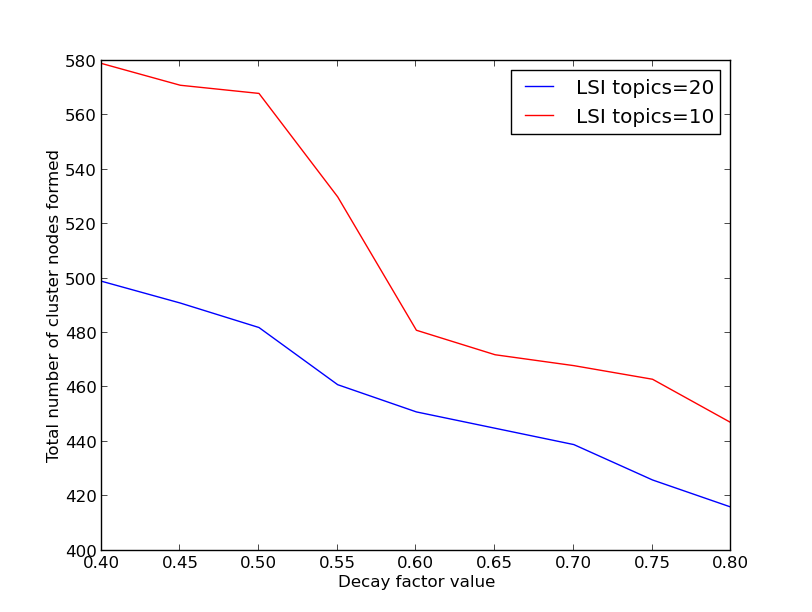}
\vspace{-0.5cm}
\hspace*{7in}\caption{Number of clusters formed vs. decay factor}
\label{fig:fig1}

\end{figure}

	Consider a run of our algorithm with decay ($\beta$) = 0.5 and number of LSI topics set to 20. The total number of clusters formed were 563(not including the individual documents at level 0), and the level-wise breakup of the number of nodes (from root to leaves) was 1-9-23-108-422. On an average, number of children of any internal node cluster was around 3-5. Figure \ref{fig:fig1} shows the number of clusters formed as a function of the decay value, keeping the number of topics constant at 20 and 10. 
	As can be seen from the graph, the number of clusters decreases steeply as $\beta$ increases. In both the plots, it can be seen that the fall in the number of clusters is highest around $\beta = 0.5$. This may indicate a possible point of interest, denoting a `dip' in the information learn.
\begin{figure}
\centering
\includegraphics[width=9.8cm]{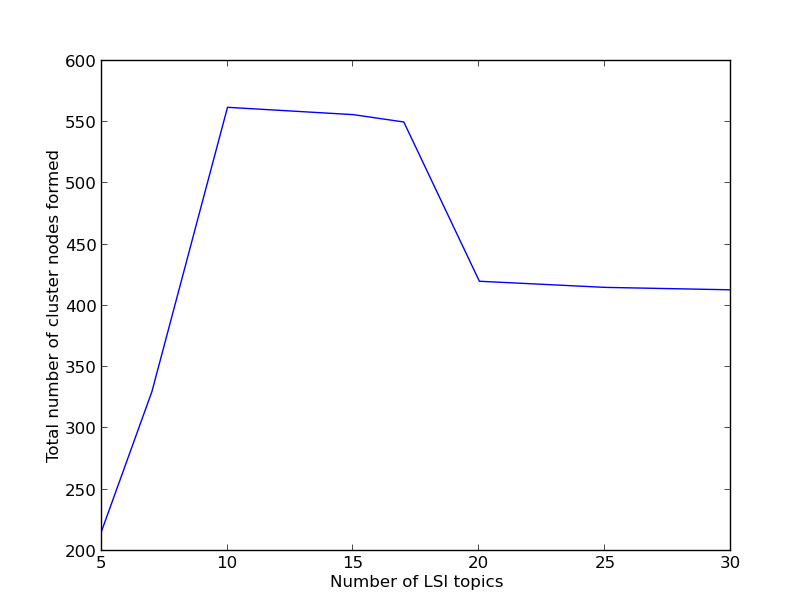}
\vspace{-0.5cm}
\centering\caption{Number of clusters formed vs. Number of LSI topics}
\label{fig:fig2}
\end{figure}
	 Figure \ref{fig:fig2} shows the number of clusters formed as a function of the number of LSI topics, keeping the decay value constant at $\beta=0.5$. It can be seen that the number of clusters increase with the number of topics upto a certain value ($\sim$10), after which they start reducing. This can be interpreted as a `gain' in information upto a certain number of topics, after which the quality of information may start reducing. Hence, we may interpret the combination of (LSI topics=10, $\beta$=0.5) to be a `special' value set for the said parameters. This will be verified soon. Over various runs of the algorithm with different constant decay values, the trends between the number of clusters and number of LSI topics remains similar. The only change occurs in the steepness of the rise and fall, which increases as the decay factor reduces.
	\begin{figure}
	\centering
	\includegraphics[width=4.6in]{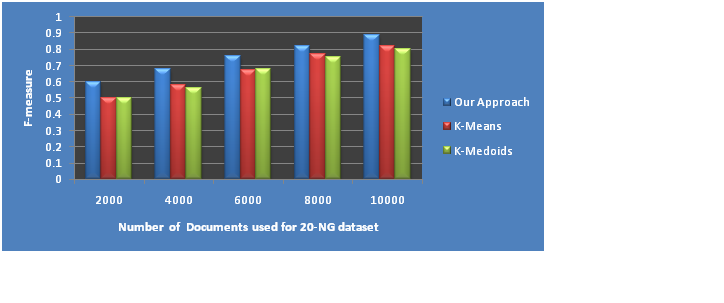}
	\vspace{-1cm}
	\centering\caption{F-measure comparisons using 20-NG Dataset}
	\label{fig:fig3}
	\end{figure} 

	We then proceed to the testing of our categorization algorithm. The `information needs' were represented by the 7535 documents not used for generating the original index. Each of the documents in the testing set was categorized to the appropriate cluster in the framework, using the algorithm described in section III(\textit{C}). Consider any document d out of the training set. Consider document d' to be the document out of the original training set, with whom d has the maximum cosine similarity. d' is said to be classified correctly, if it is categorized into a cluster having document d'. The average accuracy of the framework for a (number of LSI topics, decay) pair can be defined as the percentage of documents from the test dataset, getting classified `correctly'. Over various combinations of (number of LSI topics, decay), the average accuracy was found to be 96.46\%, with a standard deviation of 1.3\%. This shows that the categorization algorithm is pretty robust to the structure of the cluster hierarchy. The highest value of accuracy ($\sim$98.2\%) was observed for the aforementioned combination of (LSI topics= 10, $\beta$=0.5), consolidating our belief in the information portrayed by the shown graphs. The high accuracy measures the strength and quality of the clusters formed by the proposed approach. Accurate interpretation of the shown diagrams and the trends with respect to $\beta$ and number of LSI topics is ongoing. The F-measure \cite{rosell2004comparing} which is a harmonic mean of precision(fraction of a cluster that contains the documents of a specified class) and recall(to what extent, a cluster contains all documents of a specified class) has been used on the generated clusters of different sizes. The results have been compared to measure the system performance of the proposed approach with the traditional K-Means and K-Medoids algorithms and it depicted in Figure \ref{fig:fig3}. 
	\begin{displaymath}
					         F\mbox{-}measure = \frac{(2 * Precision * Recall)}{(Precision+Recall)}
	\end{displaymath}
	 
	Results show that the proposed approach outperforms the other two clustering algorithms.  
	
\section{Conclusions and Future Work}

In this paper, we presented a conceptually sound and robust framework for hierarchical clustering of documents. As we mentioned earlier, the output cluster tree can be used as a document index for an automatic retrieval system. Moreover, if every cluster is labeled and/or summarized based on its Document-Set, then navigation of the index by a human operator would be quite efficient as well. An industrial application of this work could be for domains like libraries or lawyer offices, where large amounts of text data needs to be navigated on a regular basis for information. The efficient nature of the complete algorithm ensures that the run-time of the document index generation is pretty reasonable. For example, the Python-based setup of our framework is able to create a cluster hierarchy of the 20 Newsgroups training documents in an average time of 3-4 minutes. Finally, the F-measure  over the two traditional clustering algorithms(K-Means and K-Medoids) proved the effectiveness  of the proposed approach in a better manner.

	Future work on this idea would involve making the entire cluster hierarchy dynamic. In many situations, newer and newer information in the form of documents would need to be added to an existing index of documents. In such a scenario, the hierarchy must be able to adapt itself to the changing patterns in the data by changing its own structure. This would involve creation of new nodes in the tree, merging/splitting of clusters, etc. However, the categorization algorithm would remain the same irrespective of the nature of the index. Due to the promising nature of the experimental results, we believe that further research is needed for more profound industrial applications of our current work.

\bibliographystyle{IEEEtran}
\bibliography{bibfile}

\begin{thebibliography}{10}
\providecommand{\url}[1]{#1}
\csname url@samestyle\endcsname
\providecommand{\newblock}{\relax}
\providecommand{\bibinfo}[2]{#2}
\providecommand{\BIBentrySTDinterwordspacing}{\spaceskip=0pt\relax}
\providecommand{\BIBentryALTinterwordstretchfactor}{4}
\providecommand{\BIBentryALTinterwordspacing}{\spaceskip=\fontdimen2\font plus
\BIBentryALTinterwordstretchfactor\fontdimen3\font minus
  \fontdimen4\font\relax}
\providecommand{\BIBforeignlanguage}[2]{{%
\expandafter\ifx\csname l@#1\endcsname\relax
\typeout{** WARNING: IEEEtran.bst: No hyphenation pattern has been}%
\typeout{** loaded for the language `#1'. Using the pattern for}%
\typeout{** the default language instead.}%
\else
\language=\csname l@#1\endcsname
\fi
#2}}
\providecommand{\BIBdecl}{\relax}
\BIBdecl

\bibitem{aizawa2003information}
A.~Aizawa, ``An information-theoretic perspective of tf--idf measures,''
  \emph{Information Processing \& Management}, vol.~39, no.~1, pp. 45--65,
  2003.

\bibitem{a5}
R.~H. S.~Deerwester, Susan~Dumais, ``Indexing by latent semantic analysis,''
  \emph{Proceedings of the 51st Annual Meeting of the American Society for
  Information Science}, vol.~25, pp. 36--40, 1988.

\bibitem{a16}
D.~Boley, ``Principal direction divisive partitioning,'' \emph{Data Mining and
  Knowledge Discovery}, vol.~2, no.~4, pp. 325--344, December 1998.

\bibitem{a1}
J.~B. MacQueen, ``Some methods for classification and analysis of multivariate
  observations,'' \emph{Proceedings of 5th Berkeley Symposium on Mathematical
  Statistics and Probability}, vol.~1, pp. 281--297, 1967.

\bibitem{a2}
P.~J.~R. L.~Kaufman, ``Clustering by means of medoids,'' \emph{Statistical Data
  Analysis Based on the L1–Norm and Related Methods}, pp. 405--416, 1987.

\bibitem{a7}
C.~N. Inderjit~Dhillon, Jacob~Kogan, ``Feature selection and document
  clustering,'' \emph{Survey of Text Mining}, pp. 73--100, 2003.

\bibitem{a3}
M.~I.~J. David M.~Blei, Andrew Y.~Ng, ``Latent dirichlet allocation,''
  \emph{Journal of Machine Learning Research}, vol.~3, p. 993–1022, January
  2003.

\bibitem{a4}
T.~Hofmann, ``Learning the similarity of documents : an information-geometric
  approach to document retrieval and categorization,'' \emph{Advances in Neural
  Information Processing Systems}, vol.~12, p. 914–920, 2000.

\bibitem{a6}
S.~S. Inderjit S.~Dhillon, ``Generalized nonnegative matrix approximations with
  bregman divergences,'' \emph{Advances in Neural Information Processing
  Systems}, vol.~18, 2005.

\bibitem{a10}
J.~H. Deng~Cai, Xiaofei~He, ``Locally consistent concept factorization for
  document clustering,'' \emph{IEEE Transactions on Knowledge and Data
  Engineering}, vol.~23, no.~6, pp. 902--913, June 2011.

\bibitem{a15}
W.~Song and S.~C. Park, ``Analysis of web clustering based on genetic algorithm
  with latent semantic technology,'' \emph{Proc. of Sixth International
  Conference on Advanced Language Processing and Web Information Technology},
  2007.

\bibitem{a8}
M.~E.Hasanzadeh and H.~Alinejad, ``Text clustering on latent semantic indexing
  with particle swarm optimization (pso) algorithm,'' \emph{International
  Journal of the Physical Sciences}, vol.~7, pp. 116--120, January 2012.

\bibitem{a9}
M.~Alam and K.~Sadaf, ``Web search result clustering using heuristic search and
  latent semantic indexing,'' \emph{IJCA}, vol.~44, no.~15, pp. 116--120, 2012.

\bibitem{b2}
S.~Tan and Kumar, \emph{Introduction to Data Mining}.\hskip 1em plus 0.5em
  minus 0.4em\relax Pearson Education, 2006.

\bibitem{a11}
M.~A. Nikos~Hourdakis, ``Hierarchical clustering in medical document
  collections: the bic- means method,'' \emph{JDIM}, vol.~8, pp. 71--77, 2010.

\bibitem{a14}
M.~E. Benjamin C.M.~Fung, Ke~Wang, ``Hierarchical document clustering using
  frequent itemsets,'' \emph{Proc. Of SIAM International Conference On Data
  Mining 2003}, 2003.

\bibitem{a13}
P.~C. Mahalanobis, ``On the generalised distance in statistics,''
  \emph{Proceedings of the National Institute of Sciences of India}, vol.~2,
  no.~1, pp. 49--55, 1936.

\bibitem{rosell2004comparing}
M.~Rosell, V.~Kann, and J.-E. Litton, ``Comparing comparisons: Document
  clustering evaluation using two manual classifications,'' \emph{ICON},
  vol.~4, 2004.

\end{thebibliography}

\end{document}